\documentclass[preprint,preprintnumbers,amsmath,amssymb,,nofootinbib,12pt]{revtex4-2}
\usepackage{amsfonts}
\usepackage{amssymb}
\usepackage{latexsym}
\usepackage{graphicx}
\usepackage{verbatim}
\usepackage{rotating}
\usepackage{multirow}
\usepackage{booktabs}
\usepackage{blindtext}
\usepackage[colorlinks,allcolors=blue]{hyperref}
\usepackage[usenames,dvipsnames]{color}
\usepackage[active]{srcltx}

\newcommand{\al}{\alpha}
\newcommand{\pa}{\partial}

\newcommand{\si}{\sigma}

\newcommand{\om}{\omega}

\newcommand{\rar}{\rightarrow}

\begin{document}

\title{SUSY Quantum Mechanics, (non)-Analyticity and $\ldots$ Phase Transitions}
\author{Alexander~V.~Turbiner}
\address{$^1$Instituto de Ciencias Nucleares, Universidad Nacional
Aut\'onoma de M\'exico, Apartado Postal 70-543, 04510 M\'exico,
D.F., Mexico}
\email{turbiner@nucleares.unam.mx}


\begin{abstract}
It is shown by analyzing the $1D$ Schr\"odinger equation that discontinuities in the coupling constant can occur
in both the energies and the eigenfunctions. Surprisingly, those discontinuities, which are present in the energies {\it versus} the coupling constant, are of three types only:
(i) discontinuous energies (similar to 1st order phase transitions), (ii) discontinuous first derivative in the energy while the energy is continuous (similar to 2nd order phase transitions), (iii) the energy and all its derivatives are continuous but the functions are different below and above the point of discontinuity (similar to infinite order phase transitions). Supersymmetric (SUSY) Quantum Mechanics provides a convenient framework to study this phenomenon.
\end{abstract}

\maketitle

\vspace{2pc}

\section{Introduction}

One of the general beliefs in finite-dimensional quantum mechanics is
that we deal with analytic functions of the coupling constant,
contrary to Quantum Field Theory (QFT) and/or statistical mechanics where non-analyticity/discontinuity
in temperature can occur in the form of 1st, 2nd, 3rd \ldots infinite (Berezinsky-Kosterlitz-Thouless)
order phase transitions. In this Talk we demonstrate the existence of discontinuities in coupling
constant in $1D$ quantum mechanics meaning that the functions of the coupling constant can be piece-wise analytic.
Note that a possible loss of analyticity was mentioned by the author a long time ago \cite{Turbiner:1992}
in relation to a spontaneous SUSY breaking. For simplicity we limit our consideration to shape-invariant potentials.

\bigskip

$\bullet$\ {\large \it Harmonic oscillator (example)}

\begin{figure}[hbt]
\begin{center}
    \includegraphics[width=0.5\textwidth ,angle=0]{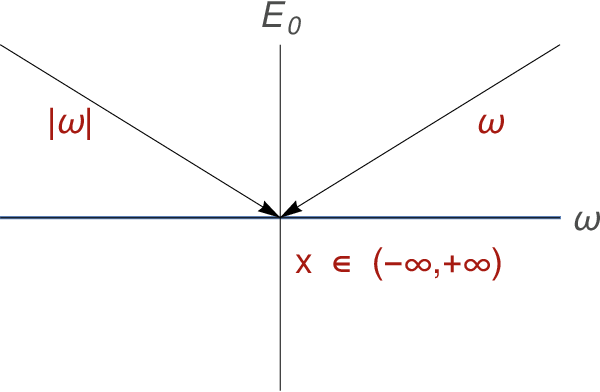}
\caption{\label{figA} Ground state energy of the harmonic oscillator {\it vs.} frequency $\om$.}
\end{center}
\end{figure}

\begin{figure}[hbt]
\begin{center}
    \includegraphics[width=0.5\textwidth ,angle=0]{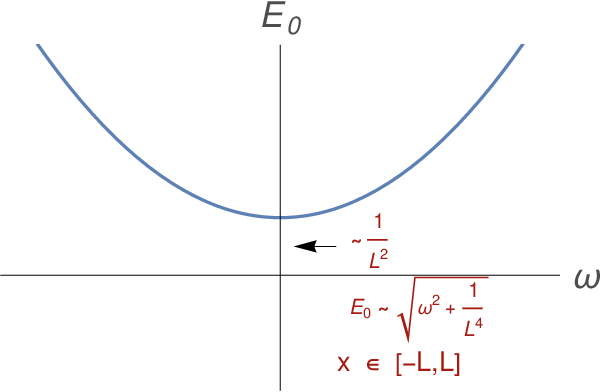}
\caption{\label{figB} Ground state energy of the harmonic oscillator in the box of size $2L$ {\it vs.} frequency $\om$: an approximate expression.}
\end{center}
\end{figure}

\bigskip
\noindent
Take the potential
\[
     V\ =\ \om^2 x^2\ ,
\]
of one-dimensional harmonic oscillator and construct the Hamiltonian
\begin{equation}
\label{H-ho}
   H\ =\ -\, \pa_x^2\ + \ \om^2 x^2 \ ,\ x \in (-\infty\ , +\infty)\ ,
\end{equation}
in atomic units with mass $m=1/2$, where  $\om$ plays the role of frequency.
It is evident that the ground state energy
\[
   E_0\ =\ \om \quad \mbox{for} \ \om > 0\quad , \quad E_0\ =\ |\om| \quad \mbox{for} \ \om < 0\ ,
\]
see Fig.\ref{figA}.
In turn, the ground state eigenfunction

\[
   \Psi_0\ =\ e^{-\om x^2/2} \quad \mbox{for} \ \om > 0\quad , \quad \Psi_0\ =\ e^{-|\om| x^2/2} \quad \mbox{for}
   \ \om < 0 \ .
\]
Needless to say both the eigenvalue and eigenfunction are non-analytic at $\om=0$, however, they admit two different analytic continuations from $\om > 0$ and $\om < 0$, respetively. Their respectful analytic continuations to $\om < 0$ or $\om > 0$, correspondingly, do exist but does not correspond to the physics reality, the eigenfunction becomes non-normalizable in these domains, respectively.
This is the simplest example of a discontinuity in the parameter $\om$ of the Hamiltonian, which itself is continuous in this parameter: there are no singularities in the eigenvalues and eigenfunctions but they are analytically disconnected at some point in the parameter space! It is evident that the analyticity in eigenstates is restored when the harmonic oscillator is placed in a finite box $x \in (-L\ , +L)$: the analytic continuation exists, see Fig.\ref{figB}. A similar situation occurs for the excited states.

\section{ SUSY Quantum Mechanics, see e.g. \cite{SUSY:1995}}

\bigskip

\noindent
Let us take SUSY Quantum Mechanics with two supercharges:
\[
     Q\ =\ \si_- \,(i \pa_x + i w(x))\ ,\quad Q^2=0 \ ,
\]
\[
     {\bar Q}\ =\ \si_+ \,(i \pa_x - i w(x))\ ,\quad {\bar Q}^2=0 \ ,
\]
where $\si_{\pm}$ are Pauli matrices, $w(x)$ is called the {\color{blue} superpotential},
and then form the Hamiltonian
\[
  {\tilde H}\ =\ \{ Q , {\bar Q}  \} \ =\ -\pa_x^2 + w^2 + \si_3 w' \ ,
\]
here $\{ a , b \}\ =\ ab + ba$ is the anticommutator, $\si_{3}$ is a diagonal
Pauli matrix, $\mbox{\it diag}(1,-1)$. Hence, ${\tilde H}$ is the $2 \times 2$ diagonal Hamiltonian matrix. The spectral problem
\[
    {\tilde H}\,\Psi\ =\ {\tilde E}\,\Psi \ ,
\]
where $\Psi, {\tilde E}$ are 2-columns, is reduced to two disconnected spectral problems

\[
  H_B \Psi_B \equiv (-\pa_x^2 + w^2 - w') \Psi_B\ =\ E_B \Psi_B \ ,
\]
\[
  H_F \Psi_F \equiv (-\pa_x^2 + w^2 + w') \Psi_F\ =\ E_F \Psi_F \ ,
\]
for the so-called bosonic and fermionic sectors, respectively. If the ground state
$\Psi^{(0)}_B = e^{-\int w dx} \in L_2(R)$, the SUSY is {\color{blue} exact} and the lowest eigenvalue of the bosonic sector vanishes, $E^{(0)}_B = 0$ (zero mode), otherwise SUSY is broken and $E^{(0)}_B \neq 0$. Let us introduce a parameter $a$ into the superpotential
\[
   w\ \rar \ a\, w \ .
\]

{\bf What are analytic properties of the spectrum wrt the parameter $a$?} \\[4pt]

It is evident that
\[
    H_B(a)\ =\ H_F(-a)
\]
thus, the bosonic and fermionic Hamiltonians are interchanged!

Let us consider two spectral problems:
\[
  H_B \Psi_B \equiv (-\pa_x^2 + a^2\,w^2 - a\,w') \Psi_B\ =\ E_B \Psi_B \ ,
\]
with spectra $E^{(0)}_B , E^{(1)}_B , E^{(2)}_B \ldots$ and
\[
  H_F \Psi_F \equiv (-\pa_x^2 + a^2\,w^2 + a\,w') \Psi_F\ =\ E_F \Psi_F \ ,
\]
with spectra $E^{(0)}_F , E^{(1)}_F , E^{(2)}_F \ldots$.
It is well known that if SUSY is exact the spectra of these two spectral problems are remarkably related
\[
   E^{(1)}_B=E^{(0)}_F \ ,\ E^{(2)}_B=E^{(1)}_F\ , \ldots \ ,
\]
hence,
\[
   E^{(1)}_B(a)\ =\ E^{(0)}_F(-a) \ .
\]
Now we consider several examples.

\bigskip

\subsection{\large $w\ =\ \om x$\ -\ Harmonic oscillator}

\bigskip

\begin{figure}[hbt]
\begin{center}
    \includegraphics[width=0.5\textwidth ,angle=0]{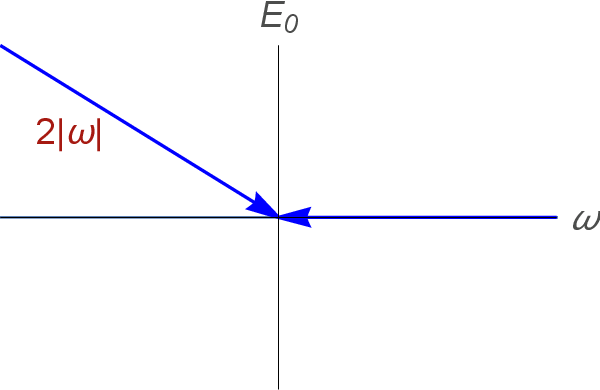}
\caption{\label{figC} The harmonic oscillator: Ground state energy $E_0$ {\it vs.} frequency $\om$}
\end{center}
\end{figure}

The potentials of the bosonic and fermionic sectors are
\[
    V_B\ =\  \om^2 x^2 - \om \ ,
\]
\[
    V_F\ =\  \om^2 x^2 + \om \ ,
\]
respectively, where $a=\om$. Hence, for $\om > 0$ the ground state energy is
\[
      E_0\ =\ 0 \ ,
\]
while for $\om < 0$ the ground state energy takes the form
\[
      E_0\ =\ 2 \,|\om| \ .
\]
It is evident that a discontinuity in the first derivative of $E_0$ occurs at $\om=0$, see Fig.\ref{figC}, this resembles a 2nd order type phase transition at $\om=0$. In a similar way discontinuity at $\om=0$ occurs for the excited states.

\bigskip

\subsection{\large  $w\ =\ a x^3$\ -\ primitive quasi-exactly-solvable (QES) sextic oscillator,
     see \cite{Turbiner:1988}}

\bigskip

\begin{figure}[hbt]
\begin{center}
    \includegraphics[width=0.4\textwidth ,angle=0]{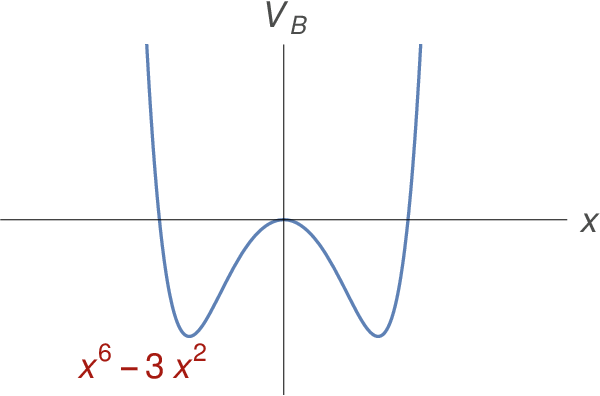}
\caption{\label{figD} Bosonic potential $V_B(x; a)$ (\ref{VB6}) at $a>0$, where the ground state energy $E_0=0$ for any $a>0$}
\end{center}
\end{figure}

\begin{figure}[hbt]
\begin{center}
    \includegraphics[width=0.4\textwidth ,angle=0]{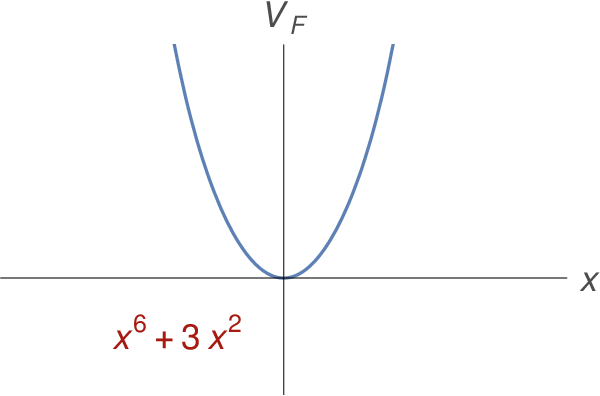}
\caption{\label{figE} Bosonic potential $V_B(x; a)$ (\ref{VB6}) at $a < 0$, where the ground state energy $E_0 \neq 0$ for any $a<0$}
\end{center}
\end{figure}

The potentials of the bosonic and fermionic sectors are
\begin{equation}
\label{VB6}
  V_B(a)\ =\  a^2 x^6 - 3 a x^2  \quad \mbox{(double well)} \ ,
\end{equation}
see Fig.\ref{figD},
\begin{equation}
\label{VF6}
    V_F(a) \ =\ V_B(-a)\ =\ a^2 x^6 + 3 a x^2   \quad \mbox{(single well)} \ ,
\end{equation}
see Fig.\ref{figE}, respectively, hence, for $a > 0 $ the ground state energy
of the bosonic sector
\[
      E_0\ =\ 0\ ,
\]
while for $a < 0$ the ground state energy of the fermionic sector
\[
      E_0\ =\ 1.935842\ |a|^{1/2}\ ,
\]
cf. \cite{Turbiner:1992}.
Thus, the discontinuity in the first derivative of $E_0$ occurs at $a=0$, see Fig.\ref{figJ} below, this resembles a 2nd order type phase transition at $a=0$. A similar discontinuity appears for the energy of
any excited state.

\bigskip

\subsection{\large $w\ =\ a x^2$\ \ -\ \ SUSY-{\color{blue} broken} quartic oscillator,
$\Psi_0 \notin L_2(R)$, $E_B^{(0)}\neq 0$}

\bigskip

The potentials of the bosonic and fermionic sectors are
\begin{equation}
\label{VB4}
    V_B\ =\  a^2 x^4 - 2 a x  \quad \mbox{(single well)}\ ,
\end{equation}
\begin{equation}
\label{VF4}
    V_F\ =\  a^2 x^4 + 2 a x   \quad \mbox{(single well)}\ .
\end{equation}
respectively. They are symmetric wrt $x \rar -x$ and/or $a \rar -a$,
\[
    V_B (x, a)\ =\ V_F (-x, a)\ =\ V_F (x, -a)\ =\ V_B (-x, -a) \ .
\]
The ground state energy
\[
      E_0(a)\ =\  0.562 136\ |a|^{2/3} \ ,
\]
is symmetric wrt $a \rar -a$ \footnote{We thank J C del Valle for the numerical calculation of $E_0(1)$ in the Lagrange Mesh method, see \cite{dV:2024}.}.

Thus, the ground state energy $E_0$ is continuous but the discontinuity in the first derivative of $E_0$
occurs at $a=0$, see Fig.\ref{figF} below, this resembles a 2nd order type phase transition at $a=0$. Similar behavior appear for the energies of the excited states.

\bigskip

\subsection{\large $w\ =\ a x^3 + b x$\ \ -\ \ generalized primitive QES sextic polynomial oscillator \cite{Turbiner:1988}}

\bigskip

Take the two-term superpotential
\[
     w\ =\ a x^3 + b x \ .
\]
The ground state function at $a > 0$
\begin{equation}
\label{psi6}
  \Psi_0\ =\ e^{-\frac{a}{4} x^4 - \frac{b}{2} x^2}\ ,
\end{equation}
is normalizable with vanishing ground state energy, $E_0 = 0$, at any $b$.
The bosonic potential takes the form
\begin{equation}
\label{VB6-gen}
  V_B\ =\  a^2 x^6 + 2 ab x^4 + (b^2 - 3 a) x^2 - b  \equiv V(x; a, b) \ ,
\end{equation}
while the fermionic potential is of the form
\begin{equation}
\label{VF6-gen}
    V_F\ =\  a^2 x^6 + 2 ab x^4 + (b^2 + 3 a) x^2 + b  \equiv V(x; -a, -b)\ ,
\end{equation}
respectively.

\bigskip
\centerline{\large \bf Remarkable Observation-I by Herbst and Simon \cite{H-Simon:1978}}
\bigskip

\begin{figure}[hbt]
\begin{center}
    \includegraphics[width=0.4\textwidth ,angle=0]{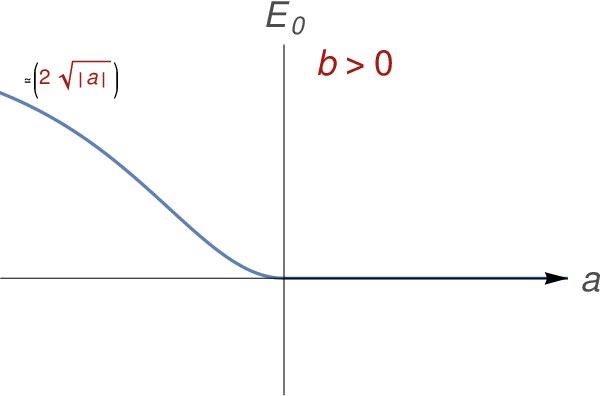}
    \caption{\label{figK} Energy of the ground state for potential (\ref{VB6-gen}) for $b > 0$ {\it vs.} $a$
    (the infinite order type phase transition)}
\end{center}
\end{figure}

\begin{figure}[hbt]
\begin{center}
    \includegraphics[width=0.4\textwidth ,angle=0]{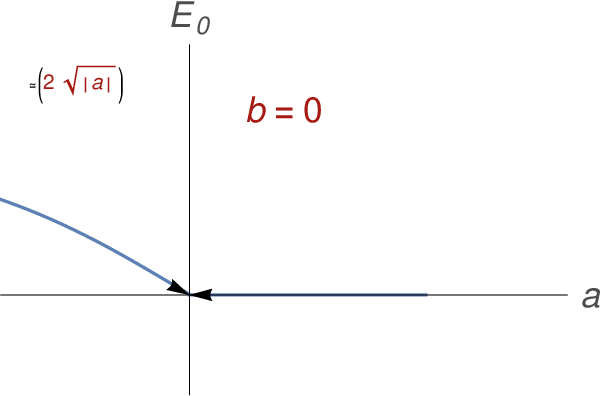}
    \caption{\label{figJ} Energy of the ground state for potential (\ref{VB6-gen}) for $b=0$ {\it vs.} $a$
    (the 2nd order type phase transition)}
\end{center}
\end{figure}

\begin{figure}[hbt]
\begin{center}
    \includegraphics[width=0.4\textwidth ,angle=0]{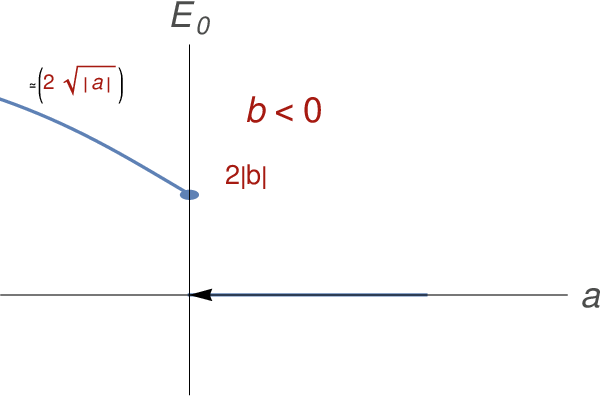}
\caption{\label{figI} Energy of the ground state for potential (\ref{VB6-gen}) for $b < 0$ {\it vs.} $a$
  (the 1st order type phase transition)}
\end{center}
\end{figure}

{\it
Take (\ref{VB6-gen})
\[
     V(x; a, b)\ \equiv\ V_B\ =\ V_0 + V_1 \ \equiv\ (b^2 x^2 - b) + (a^2 x^6 + 2 ab x^4 - 3 a x^2) \ ,
\]
and develop the perturbation theory (PT) for $V_1$ (as perturbation) by taking $V_0$ as the unperturbed potential. This PT is, in fact, in powers of the parameter $a$. For the ground state energy the PT has the form
\[
   E_0\ =\ \sum_{i=0}^{\infty} e_i\ a^i \ ,
\]
with the remarkable property that all coefficients vanish (!)
\[
     e_i=0\ , \ i=0,1,2,\ldots \ ,
\]
which was proved rigorously in \cite{H-Simon:1978}, although it is evident from the physics point of view.

If $a > 0$, the Schr\"odinger equation can be solved exactly by finding the square-integrable nodeless eigenfunction (\ref{psi6}) in the explicit form with the ground state energy $E_0=0$, however, for $a<0$\ the explicit solution (\ref{psi6}) becomes non-square-integrable, although, the square integrable ground state eigenfunction of (\ref{VF6-gen}) exists, hence, the energy $E_0$ is {\bf NOT} zero(!), see \cite{Turbiner:1992}.
}
For $b > 0$ the energy can be represented in the form of a one-instanton expansion at small (in modulo)
negative $a$,
\begin{equation}
\label{EF6}
     E_0\ =\ \beta |a|^{1/2}\ e^{-\al \frac{b}{\sqrt{|a|}}}\ (1 + \ldots)\ ,\ a \rar -0\ ,
\end{equation}
where
\[
    \beta=1.935842 \ ,\ \al=0.149\ ,
\]
cf. \cite{Turbiner:1992}. Hence, the energy is exponentially-small when $a \rar -0$. This is the remarkable example that sum of infinitely-many zeroes is {\bf not} equal to zero! Eventually, this leads to an infinite order type phase transition, see Fig.\ref{figK}. From another side if $b=0$ in (\ref{VB6-gen}): $V=V(x;a,0)$, we will arrive at a 2nd order type phase transition, see Fig.\ref{figJ}. If $b < 0$, the energy gets discontinuous at $a=0$ and the plot of the energy versus $a$ looks like a 1st order phase transition, see Fig. \ref{figI}. Similar behavior appears for the energies of the excited states in the potential $V_B$ (7) as well as for other QES potentials, see \cite{Turbiner:1988}.


\bigskip

\subsection{\large $w=a x^2 + b x$ - primitive, SUSY-broken, quartic anharmonic oscillator}

\bigskip

Take the superpotential

\[
     w\ =\ a x^2 + b x
\]
The function (a formal solution of the Schr\"odinger equation) at real $a$
\begin{equation}
\label{psi4}
  \Psi_0\ =\ e^{-\frac{a}{3} x^3 - \frac{b}{2} x^2} \ ,
\end{equation}
with (formal) energy $E_0 = 0$ at any $b$ and $a \neq 0$ is {\bf NOT} in $L_2(R)$.

Bosonic potential
\begin{equation}
\label{VB4-gen}
    V_B\ =\  a^2 x^4 + 2 ab x^3 + b^2 x^2 - 2a x - b  \equiv V(x; a, b)
\end{equation}
while the fermionic potential
\begin{equation}
\label{VF4-gen}
    V_F\ =\  a^2 x^4 + 2 ab x^3 + b^2 x^2 + 2a x + b\  =\ V(x; -a, -b)
\end{equation}

\bigskip
\centerline{\large \bf Remarkable Observation-II by Herbst and Simon \cite{H-Simon:1978}}
\bigskip

Take
\[
     V(x; a, b)\ \equiv\ V_B \ =\ V_0 + V_1 \ \equiv\ (b^2 x^2 - b) + (a^2 x^4 + 2 ab x^3 - 2 a x)
\]
and develop PT wrt $V_1$, which appears to be in powers of $a$, for the ground state energy

\[
   E_0\ =\ \sum_{i=0}^{\infty} e_i\ a^i \ .
\]
It was proved in \cite{H-Simon:1978} that all coefficients in the expansion vanish,
\[
     e_i=0\ , \ i=0,1,2,\ldots \ .
\]
Since for real $a \neq 0$ the function $\Psi_0$ (\ref{psi4}) is non-normalizable, the ground state energy $E_0 \neq 0$ and $E_0(a)=E_0(-a)$. For $b>0$ and small $a$ the energy can be written as one-instanton-type expansion
\[
   E_0\ =\ \beta |a|^{2/3}\ e^{-\al \frac{b}{|a|^{2/3}}}\ (1 + \ldots)\ ,\ a \rar \pm 0 \ ,
\]
where parameter $\al$ needs to be calculated, $\beta \sim 0.562 136$, cf..

\begin{figure}[hbt]
\begin{center}
    \includegraphics[width=0.4\textwidth ,angle=0]{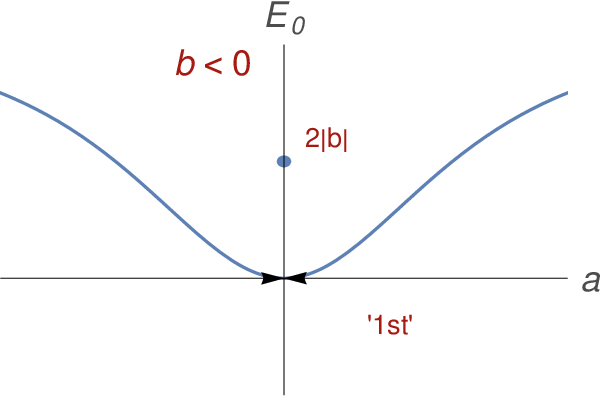}
    \caption{\label{figH} Energy of the ground state for potential (\ref{VB4-gen}) for $b<0$ {\it vs.} $a$ (the 1st order phase transition)}
\end{center}
\end{figure}

\begin{figure}[hbt]
\begin{center}
    \includegraphics[width=0.4\textwidth ,angle=0]{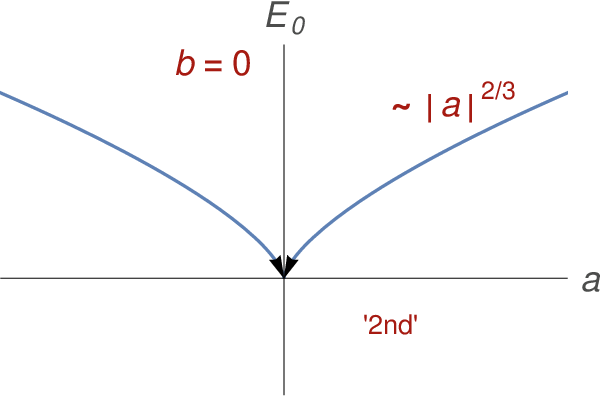}
\caption{\label{figF} Energy of the ground state for potential (\ref{VB4-gen}) for $b=0$
     {\it vs.} $a$ (the 2nd order phase transition)}
\end{center}
\end{figure}

\begin{figure}[hbt]
\begin{center}
    \includegraphics[width=0.4\textwidth ,angle=0]{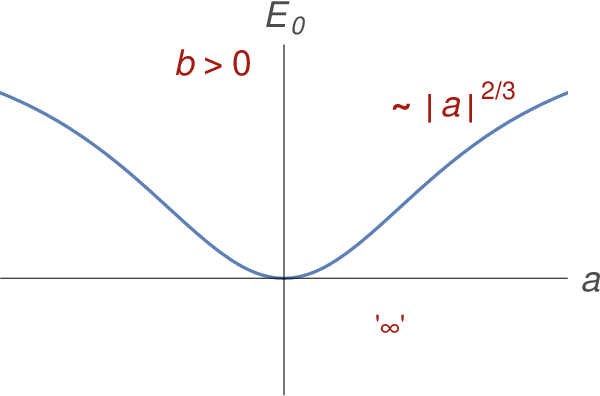}
    \caption{\label{figG} Energy of the ground state for potential (\ref{VB4-gen}) for $b>0$ {\it vs.} $a$ (the infinite order phase transition)}
\end{center}
\end{figure}

Discontinuities in $a$ of the ground state energy $E_0$, which occur at $a=0$, depending on the parameter $b$ in the potential (\ref{VB4-gen}) are presented in Figs.\ref{figH}-\ref{figG}. They are analogous to the 1st, 2nd and infinite order phase transitions, respectively.



A similar analysis to the one presented in this manuscript can be carried out for different superpotentials. All attempts to find other types of discontinuities than 1st, 2nd and infinite-order type phase transitions {\bf fail}.
Placing a system in finite box leads to disappearance of the discontinuities.


\bigskip

\section{Exact Bohr-Sommerfeld (B-S) quantization condition and SUSY quantum mechanics}

\vskip 0.4cm

Exact B-S quantization condition proposed by del Valle \& AVT \cite{BS:2021} has the form
\begin{equation}
\label{BS-exact}
     \int_{x_a}^{x_b} \sqrt{E_{exact} - V(x)}\,dx\ =\ \pi\, (N + 1/2 + \gamma (N))
\end{equation}
where $x_{a,b}(E_{exact})$ are turning points, $\gamma (N)$ is the so-called WKB correction, $\hbar=1$.
It was shown in  \cite{BS:2021} that for power-like potentials
\[
      V\ =\ |x|^m\ ,\ m > 0 \ ,
\]
the WKB correction $\gamma$ is small, dimensionless bounded function
\[
    |\gamma (N)|\ \leq \ 1/2
\]
for any $m > 0$. In particular, for the harmonic oscillator $m=2 \rar \gamma(N)=0$,
while for the quartic oscillator $m=4 \rar {\gamma}_{max}=\gamma(0) \sim 0.08$, and
for the sextic oscillator $m=6 \rar {\gamma}_{max}=\gamma(0) \sim 0.15$. Note that for the
infinite square well potential $m=\infty$ with $x \in [-1,1] \rar {\gamma}(N) = 1/2$.
The typical behavior of the WKB correction $\gamma$ is shown on Fig.\ref{m=4,6} for the case
of quartic and sextic oscillators: it is a smooth decreasing function of the quantum
number $N$ and with asymptotics $\sim 1/N$.
\begin{figure}[h]
	\centering
	\includegraphics[]{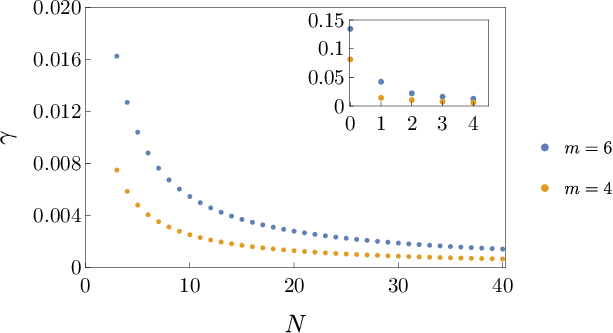}
	\caption{\label{m=4,6} WKB correction $\gamma$ {\it vs} $N$ for quartic $m=4$ and
     sextic $m=6$ oscillator potentials, see \cite{BS:2021}. Displayed points correspond to
     $N=0,1,2,3,\ldots, 40$\,. In subFigure the domain $N=0,1,2,3,4$ shown separately. }
\end{figure}

\section{ WKB correction $\gamma$ for SUSY potentials $x^{6} \mp 3 x^2$}

\bigskip

Take the potentials of the bosonic and fermionic sectors
\[
    V_{B/F}\ =\  a^2\,w^2 \mp a\,w'\ ,
\]
with energies
\[
   E^{(B)}(0)=0 \ ,\ E^{(B)}(N+1) \ =\ E^{(F)}(N)\ ,\ N=0,1,2, \ldots \ ,
\]
assuming that SUSY is exact. Exact B-S quantization condition (\ref{BS-exact})
\begin{equation}
\label{BS-exact-w}
   \int_{x_a}^{x_b} \sqrt{E^{B/F}_{exact} - a^2\,w^2 \mp a\,w'}\ dx\ =\
   \pi\,(N + 1/2 + \gamma_{B/F} (N))
\end{equation}
Take the superpotential $w=a x^3$ as an example and substitute it into (\ref{BS-exact-w}),
\begin{equation}
\label{BS-exact-lhs}
   \int_{x_a}^{x_b} \sqrt{{\tilde E}^{B/F}_{exact}\,a^{1/2} - a^2 x^6 \pm 3a x^2}\ dx \ =
\end{equation}
and change the variable
\[
   \bigg[ x\ =\ a^{-1/4}\,y \bigg]
\]
we arrive at
\[
   \int_{y_a}^{y_b} \sqrt{{\tilde E}^{B/F}_{exact}\ -\ y^6\ \pm \ 3\, y^2}\ dy \ =
\]
\bigskip
\[
   =\ \pi\,(N + 1/2 + \gamma_{B/F} (N))\ .
\]
Formally, the $a$-dependence in $\gamma_{B/F}$ disappears.

\begin{figure}[hbt]
\begin{center}
    \includegraphics[width=0.5\textwidth ,angle=0]{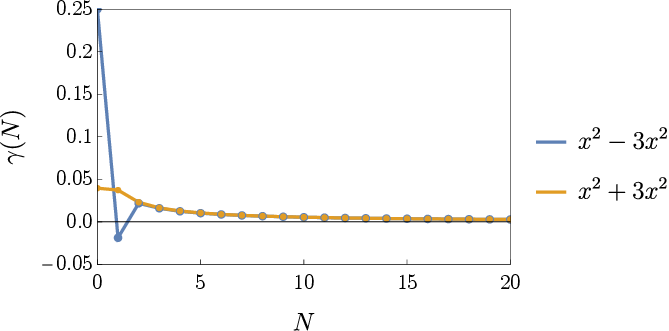}
    \caption{\label{m=6-bosonic} WKB correction $\gamma_B$ {\it vs} $N$ for the SUSY partner potentials $V_{B/F}=x^6 \pm 3 x^2$ at $N=0,1,\ldots,20$.}
\end{center}
\end{figure}

In the Lagrange Mesh method \cite{dV:2024} the Schr\"odinger equation for SUSY partner potentials
$x^{6} \mp 3 x^2$ can be easily solved for the first several eigenstates, the integral (\ref{BS-exact-lhs}) in the lhs of (\ref{BS-exact-w}) can be calculated and eventually
the WKB correction $\gamma_{B/F} (N))$ can be found. These corrections are presented
in Fig.\ref{m=6-bosonic} for the first 21 eigenstates. Non-surprisingly,
at large $N$ limit the corrections $\gamma_{B/F} (N))$ coincide with high accuracy.
%
\noindent
It must be emphasized that, in general, $\gamma^{(B)}(N+1)~\neq~\gamma^{(F)}(N)\ ,\ N=0,1,2, \ldots$ although the corresponding exact energies coincide. Those two WKB corrections are analytically disconnected in $a$.

\bigskip

\section{Conclusions}

In this Talk we give several particular examples of discontinuities in the coupling constant which occur in one-dimensional
quantum mechanics. Found discontinuities are of three types resembling the 1st, 2nd and infinite-order phase transitions
in statistical mechanics in the systems of infinite volume and infinitely many particles. By placing a system to the box leads to a disappearance of the discontinuities. What is the origin of these discontinuities - it is not clear.

\bigskip

\section*{Acknowledgements}

\bigskip

This text is dedicated to the 40 years of successful scientific career by
Professor David J Fernandez:
we sincerely congratulate David and wish him a very successful continuation.
The author thanks J C Lopez Vieyra for preparing Fig.1-10 and J C del Valle for preparing
the Fig.13 and their interest to work.
This research is partially supported by DGAPA grant IN113022 (UNAM, Mexico).

\end{document}